# Sample size calculation based on the difference in restricted mean time lost for clinical trials with competing risks

Xiang Geng[1], Zhaojin Li[1], Chengfeng Zhang[1], Yanjie Wang[1], Haoning Shen[1], Zhiheng Huang[1], Yawen Hou[2], Zheng Chen[1*]

[1]Department of Biostatistics, School of Public Health, Southern Medical University, Guangzhou, China

2 Department of Statistics and Data Science, School of Economics, Jinan University, Guangzhou, China

*Corresponding author: Zheng Chen, Email: zheng-chen@hotmail.com.

**Abstract:** Computation of sample size is important when designing clinical trials. The presence of competing risks makes the design of clinical trials with time-to-event endpoints cumbersome. A model based on the subdistribution hazard ratio (SHR) is commonly used for trials under competing risks. However, this approach has some limitations related to model assumptions and clinical interpretation. Considering such limitations, the difference in restricted mean time lost (RMTLd) is recommended as an alternative indicator. In this paper, we propose a sample size calculation method based on the RMTLd for the Weibull distribution (RMTLd$_{\text{Weibull}}$) for clinical trials, which considers experimental conditions such as equal allocation, uniform accrual, uniform loss to follow-up, and administrative censoring. Simulation results show that sample size calculation based on the RMTLd$_{\text{Weibull}}$ can generally achieve a predefined power level and maintain relative robustness. Moreover, the performance of the sample size calculation based on the RMTLd$_{\text{Weibull}}$ is similar or superior to that based on the SHR. Even if the event time does not follow the Weibull distribution, the sample size calculation based on the RMTLd$_{\text{Weibull}}$ still performs well. The results also verify the performance of the sample size calculation method based on the RMTLd$_{\text{Weibull}}$. From the perspective of the results of this study, clinical interpretation, application conditions and statistical performance, we recommend that when designing clinical trials in the presence of competing risks, the RMTLd indicator be applied for sample size calculation and subsequent effect size measurement.

**Keywords**: restricted mean time lost, sample size calculation, competing risks, design of trials

# 1. Introduction

An essential step when designing a randomized clinical trial is the calculation of the sample size to detect relevant clinical effects with sufficient significance. In trials with a time-to-event endpoint, the observed endpoint is rarely a single endpoint but multiple endpoints, and there may be competing relationships between these endpoints. For example, in published and ongoing trials for COVID-19 treatments, clinical improvement (or recovery) has been defined as an event of interest, and death is defined as a competing event[1–3]. The occurrence of the competing event (death) precludes the occurrence of the event of interest (improvement or recovery), which is a competing risk. As another example, in cardiology clinical trials, one may define nonfatal myocardial infarction or death due to cardiovascular causes as the primary endpoint of the trial and death due to other causes as a competing event[4].

In the presence of competing risks, such as in many clinical trials for COVID-19, treating competing events as having been censored is a direct approach[5–7]. However, ignoring competing risks usually leads to an overestimate of the probability that the event of interest (recovery) will occur by some time, which may lead to biased results[8–12]. In addition to the data analysis stage, during the design stage of a clinical trial, ignoring competing risks may lead to underestimating the sample size required to obtain a specified power[10] (e.g., using the single endpoint hazard ratio (HR) to calculate the sample size). Even though switching to methods of analysis for competing risks is a remedy in the data analysis stage, the analysis results may also fail to achieve the predefined power level at this time. Therefore, when competing risks exist, a common approach to estimate the sample size in the clinical design stage is to use the subdistribution hazard ratio (SHR), which is based on the between-group effect size. However, the indicator corresponding to the this method needs to meet the proportional hazard (PH) assumption and is difficult to interpret when conducting trial design and data analysis[13–16].

Thus, as an alternative to the SHR, the difference in restricted mean time lost (RMTLd) is introduced[13–19]. RMTLd can be interpreted as the difference in the mean amount of time lost due to a specific cause during a predefined time window. Moreover, it corresponds to the difference in the area under the cumulative incidence function (CIF) curve up to a specified time point and can be used to qualify the treatment effect. RMTLd is not limited by the PH assumption, and compared with the "relative" indicator that is the SHR, the clinical interpretation of the "absolute" indicator that is the RMTLd, which is based on a time scale, can easily be understood by physicians and patients. Additionally, in terms of statistical performance, the performance of the RMTLd-based hypothesis test is similar to that of the SHR-based Gray test when the PH assumption is tenable, while the performance of the RMTLd test is even better than that of the Gray test when the PH assumption fails[16]. Therefore, in this paper, RMTLd is considered as an indicator for sample size calculation in the design of clinical trials. The formula for sample size calculation based on the RMTLd was presented by Lyu et al.[15] and Wu et al.[16]. Lyu et al.[15] estimated the variance of RMTLd based on the supremum difference test to calculate the sample size ($RMTLd_{Lyu}$), but the estimation of variance is biased and relatively conservative, which may result in waste. Wu et al.[16] estimated the variance of RMTLd based on the derivation of the martingale approximation, which is more robust, and then estimated the variance of restricted time lost to calculate the sample size ($RMTLd_{Wu}$). However, neither of them considered experimental factors such as accrual time and follow-up time when calculating the sample size, so these approaches are difficult to apply in clinical trials. In this paper, we estimate the sample size for clinical trials by estimating the variance of restricted time lost for the Weibull distribution through Monte Carlo simulations ($RMTLd_{Weibull}$) by taking into account experimental factors such as accrual time, follow-up time, uniform accrual, uniform loss to follow-up and administrative censoring. In addition, we demonstrate its performance through simulations and illustrative examples.

In this article, we provide a sample size calculation for a clinical trial using the RMTLd. The paper is organized as follows. Section 2 presents the definition and estimation of the RMTL, the formula for calculating sample size, the method of correcting the variance of restricted time lost by simulation, the specific forms of the RMTL and the variance of restricted time lost for the Weibull distribution. In Section 3, we conduct simulation studies to assess the proposed sample size calculation. In Section 4, two examples are used to illustrate the proposed methods. Section 5 provides a discussion of our research.

## 2. Methods

Assume a randomized clinical trial in which $n$ patients are randomly assigned to an experimental treatment group $E$ or a control group $C$. The time-to-event and censoring times are denoted by $T_j$ and $Cen$, respectively. For simplicity, we assume that $Cen$ is independent of $T_j$. $\tau$ is the truncation time point, also called the restricted time point.

Without loss of generality, this research only considers one event of interest ($j=1$) and one competing event ($j=2$). Let $F_j(t)$ be the CIF; the CIF based on nonparametric estimation is $\hat{F}_j(t) = \sum_{t_i \leq t} (d_{ij}/n_i)\hat{S}(t_{i-1})$, where $d_{ij}$ is the number of events of type $j$ that occur at time $t_i$, the number of individuals at risk at $t_i$ is denoted by $n_i$, and $\hat{S}(t)$ is the Kaplan–Meier estimate when all events (both $j=1$ and $j=2$) are considered[15], that is, the event-free survival probability.

## 2.1 Estimation of the RMTL

Let the restricted time lost $X(\tau)$ be the $\tau$ minuses the minimum of $T_j$ and $\tau$, $X(\tau) = \tau - \min(T_j, \tau)$. Based on the estimate of the CIF, we can obtain the estimate of the RMTL and the variance of restricted time lost [15,20].

The theoretical estimation formula of the RMTL is

$$\mu_j(\tau) = E[X(\tau)] = E[\tau - \min(T_j, \tau)] = \int_0^\tau F_j(t)dt \tag{1}$$

Accordingly, the theoretical estimation formula of the variance of restricted time lost is

$$\begin{aligned}\sigma_j^2(\tau) &= RSD_j^2 = Var[X(\tau)] = Var[\tau - \min(T_j, \tau)] \\ &= 2\tau\int_0^\tau F_j(t)dt - 2\int_0^\tau tF_j(t)dt - [\int_0^\tau F_j(t)dt]^2\end{aligned} \tag{2}$$

where $RSD_j$ is the restricted standard deviation.

In practical applications, based on nonparametric CIF estimation, $\hat{\mu}_j(\tau)$ and the variance of $\tau - \min(T_j, \tau)$ ($\hat{\sigma}_j^2(\tau)$) can be obtained by using formulas (1) and (2).

## 2.2 Sample size for RMTLd

Assume there are $n_E$ patients in the experimental group and $n_C$ patients in the control group. The total sample size for the trial is $n = n_E + n_C$. The ratio of patients assigned to each group is $r = n_E/n_C$. Suppose theoretical estimates of the means and variances of $\tau - \min(T_j, \tau)$ in the experimental group and control group are $\mu_{Ej}(\tau)$ and $\sigma_{Ej}^2(\tau)$ as well as $\mu_{Cj}(\tau)$ and $\sigma_{Cj}^2(\tau)$, respectively[21]. The null hypothesis is $H_0: \mu_{Ej}(\tau) = \mu_{Cj}(\tau)$, with $\sigma_{Ej}^2(\tau)$ and $\sigma_{Cj}^2(\tau)$ unspecified. The alternative hypothesis is $H_1: \mu_{Ej}(\tau) \neq \mu_{Cj}(\tau)$. We test the null hypothesis with power $1-\beta$ at a two-sided significance level $\alpha$. Let $\Delta_j = 0$ and $\Delta_j = \mu_{Ej}(\tau) - \mu_{Cj}(\tau) \neq 0$ be the RMTLd between groups under $H_0$ and $H_1$, respectively.

Under the alternative hypothesis $H_1$, we can obtain

$$1 - \beta = \Phi[(\frac{\Delta_j}{\sqrt{\sigma_{Ej}^2(\tau)/n_{Ej}(\tau) + \sigma_{Cj}^2(\tau)/n_{Cj}(\tau)}}) - Z_{1-\alpha/2}] \quad (3)$$

$Z_p = \Phi^{-1}(p)$ is the inverse standard normal distribution function evaluated at probability $p$. The theoretical formula for the required sample size of the control group is

$$n_C = \frac{(Z_{1-\beta} + Z_{1-\alpha/2})^2}{(\Delta_j^2)/(\sigma_{Cj}^2(\tau) + \sigma_{Ej}^2(\tau)r^{-1})} \quad (4)$$

Hence, the total sample size is

$$n = n_E + n_C = \frac{(1+r)(Z_{1-\beta} + Z_{1-\alpha/2})^2}{(\Delta_j^2)/(\sigma_{Cj}^2(\tau) + \sigma_{Ej}^2(\tau)r^{-1})} \quad (5)$$

An approximate (typically conservative) sample size estimate[21], assuming that $\sigma_{Ej}^2(\tau) \approx \sigma_{Cj}^2(\tau) = \sigma_j^2(\tau)$, is given by

$$n \approx (1+r)(1+r^{-1}) = \frac{(Z_{1-\beta} + Z_{1-\alpha/2})^2}{(\Delta_j^2)/\sigma_j^2(\tau)} \quad (6)$$

We see that $n$ is minimized when $r = 1$ (equal allocation) and that $n$ increases substantially for larger or smaller $r$ (unequal allocation).

In practice, the estimates of the RMTLd and its variance are

$$\hat{\Delta}_j = \hat{\mu}_{Ej}(\tau) - \hat{\mu}_{Cj}(\tau) \quad (7)$$

$$Var(\hat{\Delta}_j) = [SE(\hat{\Delta}_j)]^2 = \hat{\sigma}_{Ej}^2(\tau)/n_E + \hat{\sigma}_{Cj}^2(\tau)/n_C \quad (8)$$

Under the null hypothesis $H_0$, the RMTLd test statistic can be computed as $Z = \frac{\hat{\Delta}_j}{\sqrt{Var(\hat{\Delta}_j)}} \sim N(0,1)$, which asymptotically follows a standard normal distribution. Therefore, hypothesis testing can be performed.

In fact, Calculating the actual required sample size needs to estimate the variance of restricted time lost such as $\sigma_{Ej}^2(\tau)$. Wu et al.[16] estimated the variance of restricted time lost by using the formula $\sigma_{Ej}^2(\tau) = n_E^* Var^*(\hat{\mu}_{Ej}(\tau))$, where $n_E^*$ and $Var^*(\hat{\mu}_{Ej}(\tau))$ can be obtained through a pilot study or previous study; however, when there is censoring before $\tau$, $n_E^* Var^*(\hat{\mu}_{Ej}(\tau))$ is no longer an accurate estimate of $\sigma_{Ej}^2(\tau)$ [21].

This paper employs Monte Carlo simulation to modify the variance of restricted time lost for the Weibull distribution and applies the scaling factor $\phi$ proposed by Royston et al.[21] to address the inflation of the variance of restricted time lost caused by censoring and obtain the corrected variance of restricted time lost $\sigma_{j,corrected}^2$.

## 2.3 Correction of the variance of restricted time lost

Assume the set of samples $T_1,...,T_m$ is independent and identically distributed from some distribution with no censoring before $\tau$ (there could be censoring after $\tau$). The RMTL may be estimated as the sample mean $\hat{\mu}_j(\tau)$, and its standard error can be expressed as $SE(\hat{\mu}_j(\tau)) = RSD_j/\sqrt{m}$.

With the censoring of some observations before $\tau$, we no longer expect $RSD_j/\sqrt{m}$ to be an accurate estimate of $SE(\hat{\mu}_j(\tau))$ since it does not reflect the increased uncertainty associated with censoring[21].

Consider a restricted sample $X_1,...,X_m$ with censoring before restricted time point $\tau$. (Note that $m$ bears no relation to the number of patients required in a trial[21].) Write $SE(\hat{\mu}_j(\tau)) = \phi \dfrac{RSD_j}{\sqrt{m}}$, where $\phi$ is a positive but unknown scaling factor proposed by Royston et al. and $\phi$ must be estimated under a known (hypothesized) Weibull distribution by using Monte Carlo simulation.

In Section 2.4, we derive the specific forms of the RMTL $\mu_{j,Weibull}$ and the variance of restricted time lost $\sigma_{j,Weibull}^2$ for the Weibull distribution and then apply $\sigma_{j,Weibull}^2$ to a Monte Carlo simulation. First, we draw a large random sample of $m$ time-to-event observations from the Weibull distribution and determine $\phi = \dfrac{\sqrt{m} SE(\hat{\mu}_j(\tau))}{RSD_{j,Weibull}}$, where $SE(\hat{\mu}_j(\tau))$ is estimated based on the derivation of the martingale approximation. Note that $RSD_{j,Weibull}$ is a known function (presented in Section 2.4) and does not need to be estimated by simulation. Given estimates of $\phi$, we can determine $\sigma_{j,corrected}^2 = (\phi RSD_{j,Weibull})^2$, and then substituting the calculated $\sigma_{j,corrected}^2$ into the sample size formula $n = \frac{(1+r)(Z_{1-\beta} + Z_{1-\alpha/2})^2}{(\Delta_j^2)/(\sigma_{Cj,corrected}^2 + \sigma_{Ej,corrected}^2 r^{-1})}$, the required sample size can be obtained.

## 2.4 RMTL for the Weibull distribution

To realize the abovementioned process of mitigating the influence of censoring by Monte Carlo simulation, we choose to assume that the survival time follows a Weibull distribution because it is flexible and can accommodate increasing, constant, and decreasing hazards[10]. In addition, the sample size calculation formula based on the Weibull distribution is relatively robust[22,23]. Therefore, this paper assumes that the survival time of the two groups of the event of interest follows the Weibull distribution and that the survival time of the two groups of the competing event also follows the Weibull distribution, and we obtain the following:

The cause-specific hazard function for the Weibull distribution is

$$\lambda_{j,Weibull}(t) = k_j \rho_j^{k_j} t^{k_j - 1} \tag{9}$$

where $k$ is a shape parameter and $\frac{1}{\rho}$ is a scale parameter.

The survival function for the Weibull distribution is

$$S_{j,Weibull}(t) = \exp\{-(\rho_j t)^{k_j}\} \tag{10}$$

The form of the cumulative incidence function for the Weibull distribution is

$$F_{j,Weibull}(t) = \int_0^t \exp\{-(\rho_1 u)^{k_1} - (\rho_2 u)^{k_2}\} k_j \rho_j^{k_j} u^{k_j - 1} du \tag{11}$$

The formulas for the RMTL $\mu_j$ and the variance of restricted time lost $\sigma_j^2$ for the Weibull distribution are:

$$\mu_{j,Weibull} = \frac{\rho_j^k}{\rho_1^k + \rho_2^k} \tau - \frac{\rho_j^k}{(\rho_1^k + \rho_2^k)^{1+\frac{1}{k}}} [\Gamma(\frac{1}{k}, 0) - \Gamma(\frac{1}{k}, ((\rho_1^k + \rho_2^k) \cdot \tau^k))] \tag{12}$$

$$\sigma_{j,Weibull}^2 = \tau^2 - (\mu - \tau)^2 - \frac{2\rho_j^k}{\rho_1^k + \rho_2^k} \{\frac{1}{2}\tau^2 + \frac{\tau}{k(\rho_1^k + \rho_2^k)^{\frac{2}{k}}} \cdot [\Gamma(\frac{2}{k}, 0) - \Gamma(\frac{2}{k}, (\rho_1^k + \rho_2^k) \cdot \tau^2)]\}$$

(13)

Once the forms of the RMTL $\mu_{j,Weibull}$ and the variance of restricted time lost $\sigma_{j,Weibull}^2$ for the Weibull distribution are obtained, the steps in Section 2.3 can be implemented. Combining them with the factor $\phi$ found by Monte Carlo simulation, $\sigma^2_{j,corrected} = (\phi RSD_{j,Weibull})^2$ can be obtained ( $RSD_{j,Weibull} = \sqrt{\sigma^2_{j,Weibull}}$ ), which is the corrected variance of restricted time lost. Then, the sample size can be calculated.

## 3、Simulations

### 3.1 Simulation setup

Next, we assessed the performance of the sample size calculation method based on RMTLd$_{Weibull}$ for the Weibull distribution data with simulations and compared it with the sample size calculation methods based on the single endpoint HR, competing risks SHR, RMTLd$_{Lyu}$ and RMTLd$_{Wu}$. Assuming a randomized controlled trial with an equal allocation in the experimental group and control group. Considering the case of one event of interest and one competing event, the accrual and loss to follow-up of the two groups were described by uniform distributions. We set the accrual period as $t_a = 18$ and the follow-up period as $t_f = 28$. The restricted time point $\tau$ was the minimum of the maximum follow-up time of the two groups[16,24]. At the same time, the sample size calculation based on the single endpoint HR and competing risk SHR also used information up to the time point $\tau$. We assumed that the survival times of the event of interest of the two groups followed the Weibull distribution and that the survival times of the two groups of the competing event also followed the Weibull distribution. We then simulated and calculated the sample size based on each method and its corresponding power in two scenarios and four censoring rates. The comparison was performed with a two-sided level $\alpha = 0.05$ and power $1 - \beta = 0.8$. The CIF curves in the two scenarios are shown in Figure 1A-1B. In Scenario A, the PH assumption is satisfied, and in Scenario B, the PH assumption is violated. The parameter settings for the simulation to generate Weibull data and the results of the sample size simulations with the number of iterations of the simulations set to 1000 are shown in Supplementary File Table S1 and Table S2.

### 3.2 Simulation results

The simulation results for the Weibull distribution data are shown in Table 1. Taking 5% censoring in Scenario A as an example, *N*=167, 232, 216, 230, and 216 are the sample sizes based on the single endpoint HR, competing risks SHR, RMTLd$_{Weibull}$, RMTLd$_{Lyu}$, and RMTLd$_{Wu}$ methods, respectively, to achieve the predefined power level of 80%. Power$_{HR}$=0.766, Power$_{SHR}$=0.642, and Power$_{RMTLd}$=0.638 are the powers of the single endpoint HR, competing risks SHR and RMTLd methods calculated when $N=167$, respectively, and the rest of the results are similar. The simulation results show that in Scenario A, when the PH assumption is met, the power of the RMTLd$_{Weibull}$ method is similar to the power of the SHR method under each censoring rate, and both are close to the predefined level of 80%. Additionally, the sample size based on the RMTLd$_{Weibull}$ method is smaller than that based on the SHR method. In Scenario B, when the PH assumption is violated, the power of the RMTLd$_{Weibull}$ method is closer to the predefined level of 80% than the power of the SHR method under each censoring rate. In addition, in Scenario B, when the censoring rate is small, the sample size based on the RMTLd$_{Weibull}$ method is much smaller than that based on the SHR method. When compared to the same RMTLd-based method, the results show that the performance of the RMTLd$_{Weibull}$ method for Weibull distributed data is similar to or slightly better than that of the RMTLd$_{Wu}$ method. The RMTLd$_{Lyu}$ method requires a larger sample size and is relatively conservative, which may result in waste.

However, if the single endpoint HR is used to calculate the sample size in the presence of competing risks. For example, under the 5% censoring in Scenario A, the sample size based on the single endpoint HR method ($N=167$) is smaller than that based on the RMTLd$_{Weibull}$ method ($N=216$) or the SHR method ($N=232$). However, when the sample size was calculated using the single-endpoint HR method in the design stage and tested based on the SHR or the RMTLd methods in the analysis stage, the power based on the SHR was 0.642, and the power based on the RMTLd was 0.683, which were far from the predefined level of 80%.

In addition, this paper also simulates data following the Gompertz distribution and log-normal distribution in two scenarios to verify the performance of the RMTLd$_{Weibull}$ method. The CIF curve is shown in Supplementary File Figure S1, and the simulation results are shown in Tables 2 and 3. According to the results, the performance of the RMTLd$_{Weibull}$ method for Gompertz distributed data or log-normally distributed data is basically similar to that for Weibull distributed data. The RMTLd$_{Weibull}$ method still performs better even when the survival time of the event of interest in the experimental and control groups in log-normally distributed data in Scenario B does not follow the Weibull distribution (experimental group $P=0.002$, control group $P=0.025$).

To assess the impact of the selection of the restricted time point $\tau$ on the performance of the RMTLd$_{Weibull}$ method, we take 5% censoring for the Weibull distribution data as an example and calculate the sample size and power when taking different values of $\tau$ in Scenarios A and B. The results are shown in Figure 2A-2D. When the PH assumption is satisfied, the sample size based on the single endpoint HR, competing risks SHR, RMTLd$_{Weibull}$, RMTLd$_{Lyu}$ and RMTLd$_{Wu}$ methods all trend downward as $\tau$ increases in Figure 2A-2B. As $\tau$ approaches the later part of the follow-up, the sample size based on the RMTLd$_{Weibull}$, RMTLd$_{Lyu}$ and RMTLd$_{Wu}$ methods is smaller than the sample size based on the SHR. The powers of the competing risks SHR, RMTLd$_{Weibull}$, RMTLd$_{Lyu}$ and RMTLd$_{Wu}$ methods are basically stable at approximately 80% with increasing $\tau$, while the power of the single endpoint HR method decreases slightly. When the PH assumption is violated, the sample size based on the RMTLd$_{Weibull}$, RMTLd$_{Lyu}$ and RMTLd$_{Wu}$ methods in Figure 2C-2D shows a trend of first decreasing and then increasing as $\tau$ increases. When $\tau$ is taken from approximately 20 to 35, the sample size based on the RMTLd$_{Weibull}$, RMTLd$_{Lyu}$ and RMTLd$_{Wu}$ method is even smaller than the sample size based on the single-endpoint HR method. The power of the RMTLd$_{Weibull}$, RMTLd$_{Lyu}$, and RMTLd$_{Wu}$ methods is basically stable at approximately 80%, while the power of the single-endpoint HR and competing risks SHR methods increases with increasing $\tau$ and eventually stabilizes at approximately 80%. The change in sample size may be affected by the relative rates of increase between the RMTLd and its variance[25]. In this paper, we find that the growth rate of the RMTLd has a greater impact on the change in sample size. For example, in Scenario A (Figure 2E), the growth rate of RMTLd remains essentially constant as $\tau$ increases, and the sample size continues to decrease and finally reaches a stable state. In Scenario B (Figure 2F), the growth rate of RMTLd first increases and then decreases as $\tau$ increases, so the sample size first decreases and then increases accordingly.

Taking $\tau=15$ under 5% censoring as an example, this paper discusses the impact of the change in accrual time and follow-up time on the sample size and power of the single endpoint HR, competing risk SHR, RMTLd$_{Weibull}$, RMTLd$_{Lyu}$ and RMTLd$_{Wu}$ methods. The results are shown in Figure 3A-3H. Regardless of whether the PH assumption is satisfied or not, the impact of the accrual time $t_a$ on the sample size and power of the five methods changes only if $t_a$ is equal to 0. In both scenarios, as the follow-up time increases, the sample size of each method first decreases and then shows a horizontal trend. The powers of the RMTLd$_{Weibull}$, RMTLd$_{Lyu}$ and RMTLd$_{Wu}$ methods fluctuate slightly until $t_f=\tau=15$ and then stabilize, while the powers of the single endpoint HR and competing risks SHR first show upward trends and then stabilize.

# 4、Illustrative examples

## 4.1 Example 1

Adaptive COVID-19 Treatment Trial 1 (ACTT-1) was a placebo-controlled trial to assess the use of remdesivir in hospitalized patients with COVID-19 [26]. The data were reconstructed; the event of interest was defined as recovery, and the corresponding competing event was death. In the ACTT-1 trial, 541 patients were assigned to the remdesivir group, and 521 patients were assigned to the placebo group, with a censoring rate of 19.2%. Figure 4 shows the CIF curves of recovery between groups.

The black solid line and the red dotted line in Figure 4 are the CIF curves of the event of interest (recovery) in the placebo group and the remdesivir group, in which the PH assumption was violated ( $P = 0.002$ ). The results obtained from the analysis based on the single endpoint HR, competing risks SHR and RMTLd methods are shown in Table 4. Although the result based on the SHR method showed a significant difference between groups (SHR=1.305, 95% CI: 1.130, 1.506), the PH assumption was not satisfied, and the true SHR may have varied with time rather than staying constant, which made clinical interpretation difficult. The RMTLs of the placebo group and the remdesivir group were 10.859 days and 13.286 days, respectively, which can be interpreted as the following statement: over the 28-day follow-up, the patients in the placebo group had an average of 10.859 postrecovery days, while the patients in the remdesivir group had an average of 13.286 postrecovery days. In other words, by the 28th day, the patients in the placebo group and remdesivir group had recovered for an average of 10.859 days and 13.286 days, respectively. The RMTLd was 2.427 days (95% CI: 1.242, 3.612), indicating that during the 28-day period, patients in the remdesivir group recovered 2.427 days earlier on average than patients in the placebo group, demonstrating the efficacy of remdesivir[16].

At the same time, the survival time of the two groups of the event of interest in the actual data did not follow the Weibull distribution after Kolmogorov–Smirnov testing (remdesivir group: $P$=0.004, placebo group: $P$=0.034). However, according to the simulation results of data following different distributions, we could still fit the survival time of each group of the event of interest and competing event in the data with Weibull parameters. The sample size was also re-estimated for the ACTT-1 trial based on the single endpoint HR, competing risk SHR, RMTLd$_{Lyu}$, RMTLd$_{Wu}$ and RMTLd$_{Weibull}$ methods by using $RMTLd = 2.427$, $CHR = 1.326$ and $SHR = 1.305$ as parameters.

The accrual period was $t_a = 58$ days, and the follow-up period was $t_f = 28$ days[26]. The allocation ratio of patients was $r = 1$. Treatment comparison was performed using a two-sided significance level $\alpha = 0.05$ with $Power = 0.8$. The results of the calculations are shown in Table 5. It can be seen that the sample size calculated based on the RMTLd$_{Weibull}$ method is $N=434$, which is not only smaller than the sample size calculated by the single endpoint HR and competing risk SHR methods but also smaller than that of the RMTLd$_{Lyu}$ and RMTLd$_{Wu}$ methods. According to the censoring rate of the actual data, it was possible to compare the data with the simulation scenario where the survival time of the two groups of the event of interest did not follow the Weibull distribution. For example, in the simulation results under 15% censoring in Table 2, the power of the RMTLd$_{Weibull}$ method should be close to the expected 80% level. Compared to the result calculated based on the SHR method, that of the RMTLd$_{Weibull}$ method required a smaller sample size and was more cost effective, illustrating the credibility and reasonableness of the result in the example validation.

Furthermore, the sample sizes based on the single endpoint HR, competing risks SHR and RMTLd$_{Weibull}$ methods were calculated in Example 1 at different time points $\tau$ for sensitivity analysis. The results are shown in Supplementary File Table S3. As the value of $\tau$ increased, the sample size based on the RMTLd$_{Weibull}$ method gradually decreased. When $\tau \geq 18$, the sample size based on the RMTLd$_{Weibull}$ method was smaller than that based on the SHR method. Therefore, the selection of $\tau$ is an important part of the RMTLd$_{Weibull}$ method, and it can be selected as the most clinically meaningful time. For example, day 28 of follow-up is usually selected in COVID-19 treatment trials[9,26].

Here, we found that the sample size based on the RMTLd$_{Weibull}$ method was larger than that based on the SHR method when $\tau < 18$ because the effect size RMTLd may be small in the early period, resulting in a large sample size needed. According to the trend of the sample size based on the RMTLd$_{Weibull}$ method in Supplementary File Table S3, which was first decreasing and then increasing, the CIF curves are shown in Figure 4. The sample size and power ranges in Example 1 should be similar to those in Figure 2C-2D. Therefore, it was inferred that although the sample size based on the SHR method was smaller than that based on the RMTLd$_{Weibull}$ method, the power of the RMTLd$_{Weibull}$ method should be more stable than that of the SHR method.

### 4.2 Example 2

The 4D trial was a randomized, double-blind, placebo-controlled trial to evaluate the efficacy of antihyperlipidemic treatment in reducing cardiovascular mortality and the frequency of nonfatal myocardial infarction. In this trial, the event of interest was defined as the composite endpoint of nonfatal myocardial infarction and death due to cardiovascular causes, with death due to other causes being the competing event[4]. The total research period of the trial was 4 years, and the accrual period was 1.5 years and assumed to be uniform.

Since data containing only the control group were obtained, we fit the survival time of the event of interest and the competing event for the control group to the Weibull distribution. Assuming that the PH assumption was satisfied, the HR was approximately 0.73, the SHR was approximately 0.75, and the data for the experimental treatment group was simulated. We set $\alpha = 0.05$, $\beta = 0.1$, and restricted time point $\tau$ to be the minimum of the maximum follow-up time between two groups, and the sample size is shown in Table 5. The results show that the sample size based on the RMTLd$_{Weibull}$ method still performs well.

### 5. Discussion

The presence of competing risks complicates the design and statistical analysis of clinical trials with time-event endpoints. The sample size calculation based on the SHR method in the trial design stage is limited by the PH assumption. However, the RMTLd method is not affected by this assumption, and the indicator itself is easy to understand and interpret, so the sample size calculation based on the RMTLd method is given for the trial design stage. The simulation result shows that when the PH assumption is tenable, the sample size and power based on the SHR and the RMTLd$_{Weibull}$ methods are basically close. When the PH assumption is violated, the sample size based on the RMTLd$_{Weibull}$ method is smaller than that based on the SHR method. Compared to the same RMTLd-based methods, the RMTLd$_{Weibull}$ method performs similarly or slightly better than the RMTLd$_{Wu}$ method and better than the relatively conservative RMTLd$_{Lyu}$ method. When the survival time of actual data does not follow the Weibull distribution, the sample size calculation based on the RMTLd$_{Weibull}$ method still performs well. In the example validation, the sample size calculation based on the RMTLd$_{Weibull}$ method continues to perform better than that based on the SHR method. Moreover, in Example 1, the sample size calculation based on the RMTLd$_{Weibull}$ method has a clear advantage over that based on the RMTLd$_{Wu}$ and RMTLd$_{Lyu}$ methods. This is because the sample size calculation based on the RMTLd$_{Weibull}$ method comprehensively considers the accrual period and follow-up period, which affects administrative censoring in trials and corrects the effect of censoring by the factor $\phi$. Therefore, the sample size calculation based on the RMTLd$_{Weibull}$ method is more appropriate for actual trials than that based on RMTLd$_{Wu}$ and RMTLd$_{Lyu}$ methods, which only use data without considering experimental factors, and the results are more reasonable.

This paper also performs a sensitivity analysis for taking different $\tau$ values. In Figure 2A-2D, when the PH assumption is satisfied, as $\tau$ is taken to be a middle to late follow-up time point, the performance of the sample size based on the RMTLd$_{Weibull}$ method is similar to or better than that based on the SHR method. When the PH assumption is violated, the performance of the sample size based on the RMTLd$_{Weibull}$ method is overall better than that based on the SHR method. The advantage is even more pronounced when $\tau$ is taken at a later time point in the follow-up. Therefore, the selection of the time point $\tau$ directly affects the performance of the sample size based on the RMTLd$_{Weibull}$ method. At the same time, according to Figure 2E-2F, it is inferred that the change in the sample size based on RMTLd$_{Weibull}$ method is affected by the growth rate of RMTLd.

This article also discusses the impact of accrual time and follow-up time on sample size and power in designs of clinical trials. As seen from Figure 3, the impact of accrual time $t_a$ on sample size and power is present only if $t_a$ is equal to 0. As the length of follow-up increases, the sample size tends to decrease and then stabilize. The power fluctuates when $t_f < \tau$, and after $t_f = \tau$, it is stable. The impact of follow-up time on sample size and power is also related to the selection of $\tau$.

Thus, the biggest challenge with designing a trial based on the RMTLd method for sample size calculation is the selection of the restricted time point $\tau$. We suggest that $\tau$ be selected as the most clinically meaningful time for evaluating the disease, the patient population and the treatment. For example, in Example 1, day 28 of the COVID-19 follow-up was taken as the $\tau$ [9,26]. When there is no obvious clinically meaningful selection, the minimum of the maximum follow-up time of the two groups in the pilot study can be selected, as it was in the simulation study and 4D trials in Example 2. Alternatively, according to the simulation results in Figure 2A-2D and Figure 3, it appears that the power tends to stabilize at the predefined level of 80% when the sample size is minimized. Therefore, this paper suggests that the time point that minimizes the sample size can also be chosen as the $\tau$ when the actual trial is generally consistent with the design stage. Furthermore, it is important to emphasize that $\tau$ should be selected during the design phase as a clinically meaningful time point and should not be changed as trial data are accumulated. When designing a trial, we recommend selecting a $\tau$ that is of clinical interest and then choosing accrual and follow-up periods based on experience or practical considerations[25].

There are also some shortcomings in the work of this paper. When performing sample size calculations, the Weibull distribution parameters and corresponding effect sizes assumed during the design stage of the trial need to be obtained through a pilot study and cannot be converted through survival, CIF or SHR. In the process of accrual and loss to follow-up, only uniform distributions are considered, and more conditions for trials can be included in further studies.

In this paper, a sample size calculation based on the RMTLd indicator that can be applied in the design stage of clinical trials is proposed. The sample size is calculated comprehensively by considering uniform accrual, the follow-up period, administrative censoring and uniform loss to follow-up. According to the simulation results, the example validation, the clinical interpretation of the indicator, application conditions and statistical performance, we recommend that the RMTLd indicator be considered for sample size calculations and subsequent analysis when designing clinical trials in the presence of competing risks.

At the same time, the work in this paper can be extended in multiple ways. Since the RMTLd test only uses data up to the restricted time point $\tau$, there is no need to follow patients beyond $\tau$, so designing trials based on the RMTLd method can potentially save the resources from a long-term follow-up. It can be combined with interim analysis in phase III clinical trials to ensure that a trial can be terminated early if there is strong evidence that the treatment is superior early in the trial or that the treatment is unlikely to have any effect on outcomes[25]. Moreover, trials based on the RMTLd should also be similar to trials based on the difference in restricted mean survival time (RMSTd). Once all patients have been accumulated, the timing of the final analysis can be determined. For example, Royston et al.[21] proposed a method to assess the maturity of accumulating trial data under a mean survival time endpoint, but its application to the design of clinical trials based on the RMTLd under competing risks remains an area for future research.

**Funding**

This work was supported by the National Natural Science Foundation of China (grant numbers 82173622, 81903411, 81673268) and the Guangdong Basic and Applied Basic Research Foundation (grant number 2022A1515011525).

**Conflict of Interest**

The authors declare no conflict of interest.

**Data availability statement**

The data that support the findings of this study in are available from the corresponding author upon reasonable request.

**Acknowledgments**

None.

**References:**

1. Dillman A, Park JJH, Zoratti MJ, et al. Reporting and design of randomized controlled trials for COVID-19: A systematic review. *Contemporary Clinical Trials*. 2021;101:106239.
2. Goldman JD, Lye DCB, Hui DS, et al. Remdesivir for 5 or 10 Days in Patients with Severe Covid-19. *N Engl J Med*. 2020;383(19):1827-1837.
3. Li L, Zhang W, Hu Y, et al. Effect of Convalescent Plasma Therapy on Time to Clinical Improvement in Patients With Severe and Life-threatening COVID-19: A Randomized Clinical Trial. *JAMA*. 2020;324(5):460.
4. Latouche A, Porcher R. Sample size calculations in the presence of competing risks. *Statist Med*. 2007;26(30):5370-5380.
5. Cao B, Wang Y, Wen D, et al. A Trial of Lopinavir–Ritonavir in Adults Hospitalized with Severe Covid-19. *N Engl J Med*. 2020;382(19):1787-1799.
6. Wang Y, Zhang D, Du G, et al. Remdesivir in adults with severe COVID-19: a randomised, double-blind, placebo-controlled, multicentre trial. *The Lancet*. 2020;395(10236):1569-1578.
7. Beigel JH, Tomashek KM, Dodd LE. Remdesivir for the Treatment of Covid-19 — Preliminary Report. *N Engl J Med*. 2020;383(10):992-994.
8. Andersen PK, Geskus RB, de Witte T, Putter H. Competing risks in epidemiology: possibilities and pitfalls. *International Journal of Epidemiology*. 2012;41(3):861-870.
9. McCaw ZR, Tian L, Vassy JL, et al. How to Quantify and Interpret Treatment Effects in Comparative Clinical Studies of COVID-19. *Annals of Internal Medicine*. 2020;173(8):632-637.
10. Maki E. Power and sample size considerations in clinical trials with competing risk endpoints. *Pharmaceut Statist*. 2006;5(3):159-171.
11. Schuster NA, Hoogendijk EO, Kok AAL, Twisk JWR, Heymans MW. Ignoring competing events in the analysis of survival data may lead to biased results: a nonmathematical illustration of competing risk analysis. *Journal of Clinical Epidemiology*. 2020;122:42-48.
12. Koller MT, Raatz H, Steyerberg EW, Wolbers M. Competing risks and the clinical community: irrelevance or ignorance? *Statist Med*. 2012;31(11-12):1089-1097.
13. Zhao L, Tian L, Claggett B, et al. Estimating Treatment Effect With Clinical Interpretation From a Comparative Clinical Trial With an End Point Subject to Competing Risks. *JAMA Cardiol*. 2018;3(4):357.
14. McCaw ZR, Claggett BL, Tian L, et al. Practical Recommendations on Quantifying and Interpreting Treatment Effects in the Presence of Terminal Competing Risks: A Review. *JAMA Cardiol*. 2022;7(4):450.
15. Lyu J, Hou Y, Chen Z. The use of restricted mean time lost under competing risks data. *BMC Med Res Methodol*. 2020;20(1):197.
16. Wu H, Yuan H, Yang Z, Hou Y, Chen Z. Implementation of an Alternative Method for Assessing Competing Risks: Restricted Mean Time Lost. *American Journal of Epidemiology*. 2022;191(1):163-172.

17. Conner SC, Trinquart L. Estimation and modeling of the restricted mean time lost in the presence of competing risks. *Statistics in Medicine*. 2021;40(9):2177-2196.
18. Lin J, Trinquart L. Doubly-robust estimator of the difference in restricted mean times lost with competing risks data. *Stat Methods Med Res*. 2022;31(10):1881-1903.
19. Lyu J, Hou Y, Chen Z. Combined Tests Based on Restricted Mean Time Lost for Competing Risks Data. *Statistics in Biopharmaceutical Research*. 2023,15(2): 332-339.
20. Andersen PK. Decomposition of number of life years lost according to causes of death. *Statist Med*. 2013;32(30):5278-5285.
21. Royston P, Parmar MK. Restricted mean survival time: an alternative to the hazard ratio for the design and analysis of randomized trials with a time-to-event outcome. *BMC Med Res Methodol*. 2013;13(1):152.
22. Han D, Chen Z, Hou Y. Sample size for a noninferiority clinical trial with time-to-event data in the presence of competing risks. *Journal of Biopharmaceutical Statistics*. 2018;28(4):797-807.
23. Han D, Hou Y, Ou C, Chen Z. Sample size determination in superiority or non-inferiority clinical trials with time-to-event data under exponential, Weibull and Gompertz distributions. *Communications in Statistics - Simulation and Computation*. 2023;52(4):1212-1224.
24. Tian L, Jin H, Uno H, et al. On the empirical choice of the time window for restricted mean survival time. *Biometrics*. 2020;76(4):1157-1166.
25. Eaton A, Therneau T, Le-Rademacher J. Designing clinical trials with (restricted) mean survival time endpoint: Practical considerations. *Clinical Trials*. 2020;17(3):285-294.
26. Beigel JH, Tomashek KM, Dodd LE, et al. Remdesivir for the Treatment of Covid-19 — Final Report. *N Engl J Med*. 2020;383(19):1813-1826.

**Table 1** Simulation results of the sample size and power based on the single endpoint HR, competing risks SHR, RMTLd$_{Weibull}$, RMTLd$_{Lyu}$, and RMTLd$_{Wu}$ methods for Weibull distributed data in two scenarios

| Censoring rate | | Scenario A | | | | Scenario B | | | |
|---|---|---|---|---|---|---|---|---|---|
| | | N | Power$_{HR}$ | Power$_{SHR}$ | Power$_{RMTLd}$ | N | Power$_{HR}$ | Power$_{SHR}$ | Power$_{RMTLd}$ |
| 5% | HR | 167 | **0.766** | 0.642 | 0.683 | 683 | **0.791** | 0.264 | 0.684 |
| | SHR | 232 | 0.893 | **0.793** | 0.818 | 2978 | 1.000 | **0.777** | 0.998 |
| | RMTLd$_{Weibull}$ | 216 | 0.863 | 0.762 | **0.789** | 915 | 0.905 | 0.328 | **0.801** |
| | RMTLd$_{Lyu}$ | 230 | 0.868 | 0.774 | **0.805** | 970 | 0.924 | 0.353 | **0.835** |
| | RMTLd$_{Wu}$ | 216 | 0.863 | 0.762 | **0.789** | 916 | 0.905 | 0.328 | **0.801** |
| 15% | HR | 187 | **0.782** | 0.681 | 0.704 | 662 | **0.813** | 0.313 | 0.659 |
| | SHR | 245 | 0.871 | **0.789** | 0.807 | 2412 | 1.000 | **0.733** | 0.986 |
| | RMTLd$_{Weibull}$ | 234 | 0.860 | 0.772 | **0.791** | 935 | 0.907 | 0.403 | **0.794** |
| | RMTLd$_{Lyu}$ | 250 | 0.872 | 0.815 | **0.825** | 990 | 0.934 | 0.460 | **0.829** |
| | RMTLd$_{Wu}$ | 236 | 0.866 | 0.794 | **0.800** | 936 | 0.907 | 0.403 | **0.794** |
| 30% | HR | 225 | **0.782** | 0.718 | 0.723 | 608 | **0.800** | 0.390 | 0.584 |
| | SHR | 266 | 0.830 | **0.786** | 0.792 | 1688 | 0.998 | **0.765** | 0.938 |
| | RMTLd$_{Weibull}$ | 260 | 0.837 | 0.790 | **0.802** | 1003 | 0.933 | 0.541 | **0.778** |
| | RMTLd$_{Lyu}$ | 276 | 0.858 | 0.822 | **0.822** | 1062 | 0.940 | 0.578 | **0.799** |
| | RMTLd$_{Wu}$ | 260 | 0.837 | 0.790 | **0.802** | 1004 | 0.933 | 0.541 | **0.778** |
| 45% | HR | 284 | **0.780** | 0.770 | 0.779 | 540 | **0.767** | 0.448 | 0.526 |
| | SHR | 310 | 0.830 | **0.798** | 0.807 | 1149 | 0.979 | **0.789** | 0.850 |
| | RMTLd$_{Weibull}$ | 306 | 0.809 | 0.784 | **0.795** | 971 | 0.961 | 0.709 | **0.797** |
| | RMTLd$_{Lyu}$ | 324 | 0.811 | 0.780 | **0.784** | 1028 | 0.966 | 0.735 | **0.813** |
| | RMTLd$_{Wu}$ | 306 | 0.809 | 0.784 | **0.795** | 972 | 0.961 | 0.709 | **0.797** |

Note: N, the sample size based on the single endpoint HR, competing risks SHR, RMTLd$_{Weibull}$, RMTLd$_{Lyu}$, RMTLd$_{Wu}$ methods.

Note: Power$_{HR}$, the power of the single endpoint HR method; Power$_{SHR}$, the power of the SHR method; Power$_{RMTLd}$, the power of the RMTLd method.

**Table 2** Simulation results of the sample size and power based on the single endpoint HR, competing risks SHR, RMTLd$_{Weibull}$, RMTLd$_{Lyu}$, and RMTLd$_{Wu}$ methods for log-normally distributed data in two scenarios

| Censoring rate | | Scenario A | | | | Scenario B | | | |
|---|---|---|---|---|---|---|---|---|---|
| | | N | Power$_{HR}$ | Power$_{SHR}$ | Power$_{RMTLd}$ | N | Power$_{HR}$ | Power$_{SHR}$ | Power$_{RMTLd}$ |
| 5% | HR | 561 | **0.823** | 0.438 | 0.421 | 1442 | **0.801** | 0.246 | 0.416 |
| | SHR | 1472 | 0.993 | **0.805** | 0.793 | 6581 | 0.999 | **0.787** | 0.974 |
| | RMTLd$_{Weibull}$ | 1543 | 0.997 | 0.811 | **0.795** | 3353 | 0.994 | 0.519 | **0.785** |
| | RMTLd$_{Lyu}$ | 1636 | 0.996 | 0.840 | **0.829** | 3558 | 0.994 | 0.528 | **0.796** |
| | RMTLd$_{Wu}$ | 1548 | 0.994 | 0.834 | **0.821** | 3364 | 0.990 | 0.526 | **0.820** |
| 15% | HR | 595 | **0.831** | 0.441 | 0.422 | 1344 | **0.802** | 0.307 | 0.410 |
| | SHR | 1565 | 0.993 | **0.792** | 0.774 | 5049 | 1.000 | **0.805** | 0.935 |
| | RMTLd$_{Weibull}$ | 1639 | 0.994 | 0.818 | **0.796** | 3348 | 0.992 | 0.639 | **0.801** |
| | RMTLd$_{Lyu}$ | 1738 | 0.996 | 0.829 | **0.808** | 3548 | 0.996 | 0.664 | **0.829** |
| | RMTLd$_{Wu}$ | 1642 | 0.996 | 0.820 | **0.812** | 3356 | 0.991 | 0.656 | **0.802** |
| 30% | HR | 669 | **0.795** | 0.406 | 0.409 | 1153 | **0.822** | 0.392 | 0.385 |
| | SHR | 1769 | 0.995 | **0.822** | 0.833 | 3135 | 0.998 | **0.807** | 0.812 |
| | RMTLd$_{Weibull}$ | 1734 | 0.994 | 0.797 | **0.812** | 2991 | 0.992 | 0.794 | **0.797** |
| | RMTLd$_{Lyu}$ | 1838 | 0.992 | 0.823 | **0.831** | 3174 | 0.999 | 0.803 | **0.805** |
| | RMTLd$_{Wu}$ | 1738 | 0.992 | 0.768 | **0.786** | 3002 | 0.998 | 0.796 | **0.787** |
| 45% | HR | 748 | **0.793** | 0.421 | 0.448 | 949 | **0.797** | 0.502 | 0.568 |
| | SHR | 1862 | 0.995 | **0.817** | 0.837 | 1906 | 0.983 | **0.820** | 0.878 |
| | RMTLd$_{Weibull}$ | 1729 | 0.986 | 0.758 | **0.798** | 1556 | 0.944 | 0.724 | **0.789** |
| | RMTLd$_{Lyu}$ | 1834 | 0.993 | 0.784 | **0.808** | 1650 | 0.963 | 0.771 | **0.825** |
| | RMTLd$_{Wu}$ | 1734 | 0.990 | 0.770 | **0.798** | 1560 | 0.936 | 0.703 | **0.773** |

Note: N, the sample size based on the single endpoint HR, competing risks SHR, RMTLd$_{Weibull}$, RMTLd$_{Lyu}$, RMTLd$_{Wu}$ methods.

Note: Power$_{HR}$, the power of the single endpoint HR method; Power$_{SHR}$, the power of the SHR method; Power$_{RMTLd}$, the power of the RMTLd method.

**Table 3** Simulation results of the sample size and power based on the single endpoint HR, competing risks SHR, RMTLd$_{Weibull}$, RMTLd$_{Lyu}$, and RMTLd$_{Wu}$ methods for Gompertz distributed data in two scenarios

| Censoring rate | | Scenario A | | | | Scenario B | | | |
|---|---|---|---|---|---|---|---|---|---|
| | | N | Power$_{HR}$ | Power$_{SHR}$ | Power$_{RMTLd}$ | N | Power$_{HR}$ | Power$_{SHR}$ | Power$_{RMTLd}$ |
| 5% | HR | 855 | **0.775** | 0.572 | 0.633 | 1246 | **0.821** | 0.266 | 0.772 |
| | SHR | 1277 | 0.904 | **0.790** | 0.832 | 5997 | 1.000 | **0.820** | 1.000 |
| | RMTLd$_{Weibull}$ | 1178 | 0.899 | 0.769 | **0.810** | 1420 | 0.866 | 0.276 | **0.798** |
| | RMTLd$_{Lyu}$ | 1250 | 0.924 | 0.794 | **0.836** | 1508 | 0.882 | 0.295 | **0.834** |
| | RMTLd$_{Wu}$ | 1182 | 0.911 | 0.756 | **0.792** | 1426 | 0.881 | 0.303 | **0.826** |
| 15% | HR | 938 | **0.775** | 0.643 | 0.668 | 1144 | **0.827** | 0.335 | 0.693 |
| | SHR | 1318 | 0.904 | **0.790** | 0.822 | 4137 | 1.000 | **0.833** | 0.997 |
| | RMTLd$_{Weibull}$ | 1248 | 0.884 | 0.763 | **0.787** | 1524 | 0.908 | 0.444 | **0.828** |
| | RMTLd$_{Lyu}$ | 1324 | 0.914 | 0.817 | **0.851** | 1616 | 0.935 | 0.441 | **0.823** |
| | RMTLd$_{Wu}$ | 1252 | 0.897 | 0.771 | **0.792** | 1528 | 0.898 | 0.410 | **0.790** |
| 30% | HR | 1131 | **0.786** | 0.709 | 0.723 | 975 | **0.807** | 0.438 | 0.600 |
| | SHR | 1407 | 0.857 | **0.788** | 0.805 | 2427 | 0.997 | **0.826** | 0.931 |
| | RMTLd$_{Weibull}$ | 1383 | 0.857 | 0.768 | **0.781** | 1588 | 0.955 | 0.663 | **0.832** |
| | RMTLd$_{Lyu}$ | 1466 | 0.870 | 0.805 | **0.813** | 1684 | 0.964 | 0.673 | **0.844** |
| | RMTLd$_{Wu}$ | 1386 | 0.865 | 0.798 | **0.803** | 1592 | 0.950 | 0.645 | **0.803** |
| 45% | HR | 1408 | **0.768** | 0.735 | 0.744 | 803 | **0.811** | 0.581 | 0.632 |
| | SHR | 1574 | 0.798 | **0.769** | 0.775 | 1465 | 0.959 | **0.807** | 0.858 |
| | RMTLd$_{Weibull}$ | 1593 | 0.810 | 0.775 | **0.777** | 1231 | 0.934 | 0.752 | **0.814** |
| | RMTLd$_{Lyu}$ | 1688 | 0.859 | 0.810 | **0.810** | 1304 | 0.941 | 0.773 | **0.813** |
| | RMTLd$_{Wu}$ | 1596 | 0.822 | 0.783 | **0.798** | 1234 | 0.945 | 0.747 | **0.799** |

Note: N, the sample size based on the single endpoint HR, competing risks SHR, RMTLd$_{Weibull}$, RMTLd$_{Lyu}$, RMTLd$_{Wu}$ methods.

Note: Power$_{HR}$, the power of the single endpoint HR method; Power$_{SHR}$, the power of the SHR method; Power$_{RMTLd}$, the power of the RMTLd method.

**Table 4** Results of the analysis of the recovery of patients in the remdesivir and placebo groups in Example 1

| Test | $\tau$ | Remdesivir (95% CI) | Placebo (95% CI) | HR/SHR/RMTLd (95% CI) | $P$ |
|---|---|---|---|---|---|
| HR | - | \ | \ | 1.326* (1.145, 1.536) | <0.001 |
| SHR | - | \ | \ | 1.305** (1.130, 1.506) | <0.001 |
| RMTLd | 28 | 10.859 (10.017, 11.701) | 13.286 (12.452, 14.120) | 2.427# (1.242, 3.612) | <0.001 |

Note: * Results of the analysis based on the single endpoint HR method, ** Results of the analysis based on the SHR method, # Results of the analysis based on the RMTLd method

**Table 5** Re-estimation of the sample size for Example 1 and Example 2

| Method | Sample Size | |
|---|---|---|
| | Example 1 | Example 2 |
| HR | 565 | 758 |
| SHR | 639 | 910 |
| RMTLd$_{Weibull}$ | 434 | 873 |
| RMTLd$_{Lyu}$ | 548 | 926 |
| RMTLd$_{Wu}$ | 518 | 874 |

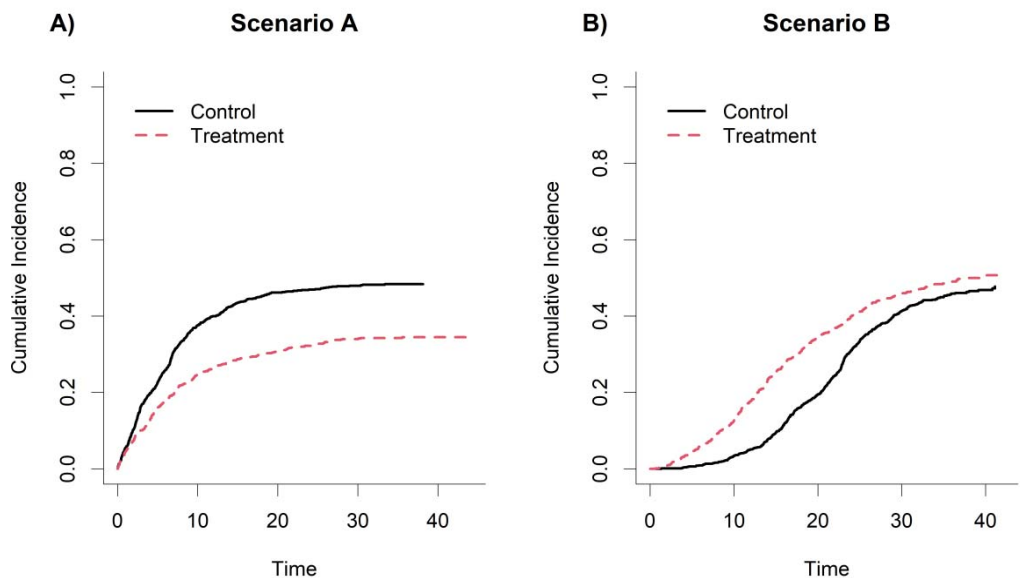

**Figure 1** CIF curves for the event of interest in two scenarios for Weibull distributed data in the simulation study

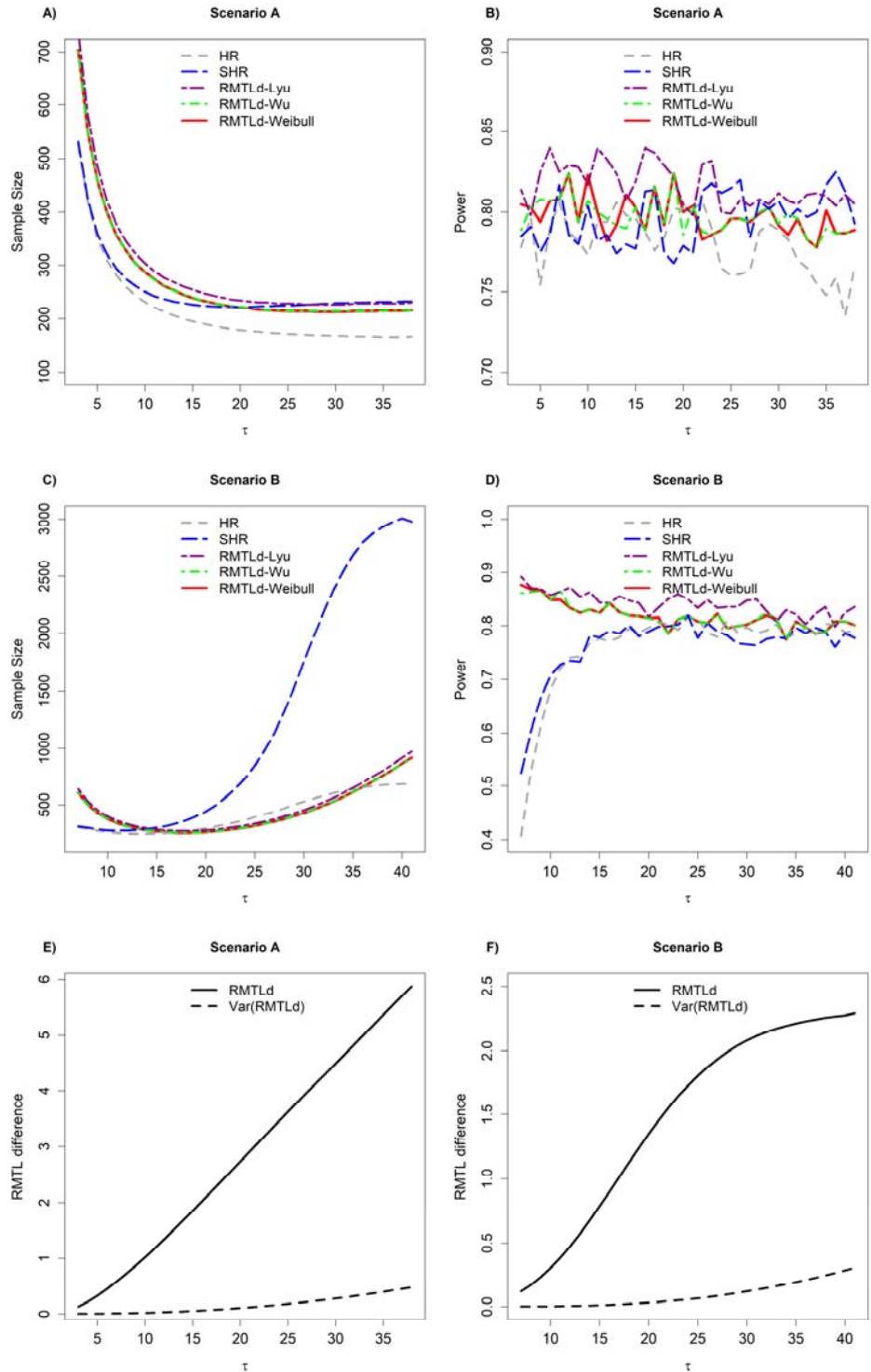

**Figure 2** Variations in sample size and power based on the single endpoint HR, competing risks SHR, RMTLd$_{Weibull}$, RMTLd$_{Lyu}$, and RMTLd$_{Wu}$ methods and variations in effect size RMTLd with its variance as $\tau$ is varied in two scenarios

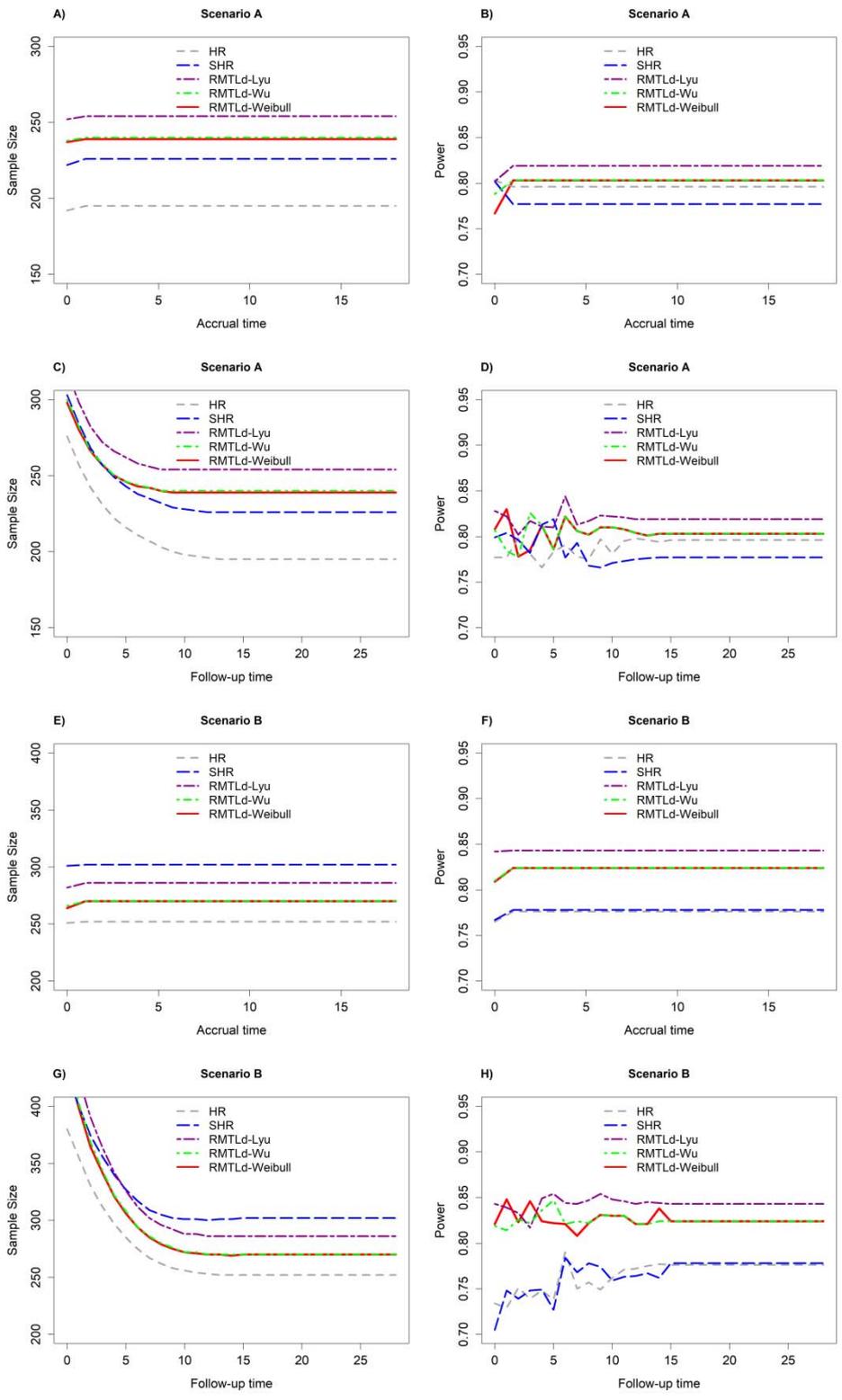

**Figure 3** Variations in sample size and power based on the single endpoint HR, competing risks SHR, RMTLd$_{Weibull}$, RMTLd$_{Lyu}$, and RMTLd$_{Wu}$ methods with changes in accrual time and follow-up time in two scenarios

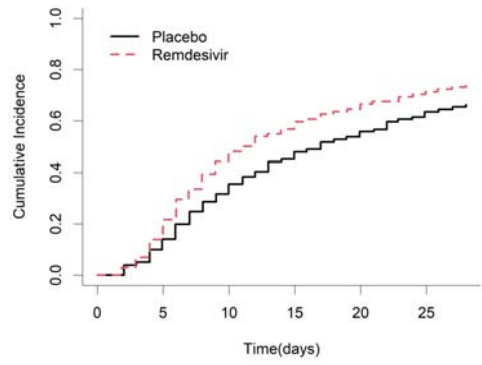

**Figure 4** CIF curves for the recovery of patients in Example 1